\documentstyle[preprint,aps]{revtex}
\draft
\begin{document}
\preprint{\today}
\title{Some Finite Size Effects in Simulations of Glass Dynamics}
\author{J\"urgen Horbach, Walter Kob\cite{wkob}, Kurt Binder}
\address{Institut f\"ur Physik, Johannes Gutenberg-Universit\"at,
Staudinger Weg 7, D-55099 Mainz, Germany}
\author{C. Austen Angell}
\address{Department of Chemistry, Arizona State University, Tempe, AZ
85287-1604}
\maketitle

\begin{abstract}
We present the results of a molecular dynamics computer simulation in
which we investigate the dynamics of silica. By considering different
system sizes, we show that in simulations of the dynamics of this strong
glass former surprisingly large finite size effects are present. In
particular we demonstrate that the relaxation times of the incoherent
intermediate scattering function and the time dependence of the mean
squared displacement are affected by such finite size effects. By
compressing the system to high densities, we transform it to a fragile
glass former and find that for that system these types of finite size 
effects are much weaker. 

\end{abstract}

\narrowtext


The dynamics of supercooled liquids and the related phenomenon of the
glass transition has been the focus of interest for a long time.  In
the last ten years a particularly intense activity could be observed in
this field, which was triggered on the one hand by new theoretical
developments, such as the so-called mode-coupling theory, and on the
other hand by the progress in experimental techniques which allowed to
make new types of experiments to investigate such systems~\cite{glass}.
However, despite all these efforts the underlying mechanism for the
dramatic (by some 15 decades) slowing down of the dynamics of the
liquid upon cooling is still a matter of debate and thus the focus of
many investigations.

In these years the dynamics of supercooled liquids has also been
investigated intensively with the help of computer
simulations~\cite{glass_simul}.  Because such simulations permit access
to the whole microscopic information of the system at any given time,
such investigations permit measurement of observables which are not
accessible in todays experiments but for which the theories make
predictions that can be tested. Therefore computer simulations are a
very useful addition to theoretical and experimental investigations in
the field of supercooled liquids and the glass transition.

With very few exceptions~\cite{boson_simul,angell95,signorini90}, the
computer simulations that investigated the {\it dynamical} properties
of supercooled liquids focused on models in which the interactions
between the particles are short-ranged.  The reason for this is that
only for such simple models was it possible to make sufficiently long
simulations to allow the system to equilibrate even at relatively low
temperatures and thus to investigate the supercooled regime. By
determining the temperature dependence of the diffusion constant or of
structural relaxation times it was found (see e.g. Ref.~\cite{kob94})
that these systems are so-called fragile glass formers~\cite{angell85}. 
The other
extreme in this classification scheme are the so-called strong 
glass formers~\cite{angell85} and the materials belonging to this class 
usually have interactions that are long ranged, such as the Coulomb
interaction in SiO$_2$. Because of the significantly increased cost in
computer time that is needed to handle such long range forces,
the dynamics of these types of systems have only been
investigated by computer simulations in a few cases. In {\it
real} experiments the dynamics of strong glass formers have been
investigated already for many years and it was found that apart from
the quantitative difference between the dynamics of strong and fragile
glass formers there is also a qualitative difference, in that the
spectra of the former show, e.g., a feature which today is called the
boson peak~\cite{bosonpeak}. Since the microscopic origin of this
feature is still unclear it would be interesting to do computer
simulations of strong glass formers in order to gain some insight into
the microscopic dynamics of such systems and some attempts in this
direction have already been made~\cite{boson_simul,angell95}.

Because of the long runs needed to equilibrate the systems at low
temperatures it is natural to use systems that are rather
small, usually between a few hundred and a thousand particles. Such small
systems have generally been thought to be large enough to avoid finite
size effects. That this is not {\it quite} the case was demonstrated by
Lewis and Wahnstr\"om in a computer simulation of the fragile
glass former orthoterphenyl~\cite{lewis94}, in that they showed 
that the intermediate scattering function has a small blip at around
5ps, the time that it takes a sound wave to cross the simulation box.
By investigating the dynamics of a larger system Lewis {\it et al.}
showed that the observed blip was indeed a finite size effect. However,
because of its smallness the authors concluded correctly that this
finite size effect can be disregarded when investigating the dynamics
of that system.

In the following we will demonstrate that although this finite size
effect can be disregarded for {\it fragile} glass formers, it cannot be
neglected in the case of {\it strong} glass formers, because it is much
more pronounced in the latter type of systems.

The potential we use is the one proposed by van Beest {\it et
al.}~\cite{beest90}, which seems to be quite reliable to describe
amorphous SiO$_2$~\cite{vollmayr96}.  The total number of particles for
our simulations, $N$, was 336, 1002, 3006 and 8016. In order to be able
to compare the dynamics of the systems with the different sizes, we
needed a well defined ensemble.  In such simulations it is common to
use the $(N,p,T)$ ensemble or the $(N,V,T)$ ensemble. However, for
the type of question investigated here, these ensembles are not very
useful, since the only way to realize them for finite systems is to
introduce an {\it artificial} dynamics (fluctuating shape of the
simulation box or coupling to a heat bath). Since our goal is to study
the finite size effects of the {\it real} dynamics of the system, we
therefore decided to fix instead the density and the total energy per
particle.

The equations of motion were integrated with the velocity form of the
Verlet algorithm with a step size of 1.6fs. The systems were first
equilibrated at a temperature of 5200K, a temperature at which the
relaxation times of silica are very short. Subsequently the velocities of
the particles were changed such that the mean total energy per particle was
$-18.0345$eV, which corresponds to a temperature of around 3760K. At
this energy the system was propagated for at least 50,000 time steps
in order to allow it to reach equilibrium. After that the production
run of 30,000 steps was started and the quantities of interest
measured. In order to improve the statistics of the results we used 20,
10, 5 and 2 independent runs for the $N=336$, 1002, 3006 and
the $N=8016$ system. 

Subsequently we changed the energy of the system to a value of
$-18.6627$eV, which corresponds to a temperature around 1600K. We let
the system propagate again for 100,000 time steps in order to allow it
to relax from the quench. However, at this low temperature the
relaxation time of the system by far exceeds this equilibration time,
thus the resulting configurations were in some way relaxed from the
quench, but not typical equilibrium configurations for that energy.

One of the natural quantities to investigate the dynamics of a fluid
like system is the incoherent intermediate scattering function
$F_{i}^{(s)}(q,t)=\langle \delta \rho_i(q,0) \delta \rho_i^{\star}(q,t)
\rangle$, where $q$ is the wave vector and $i \in
\{\mbox{Si,O}\}$. In Fig.~\ref{fig1} we show the time dependence of
this function for the oxygen particles for all system sizes
investigated. The wave-vector $q$ is 2.8\AA$^{-1}$, the location of the
maximum in the structure factor of the O-O pairs. From the main figure
we see that for short times, i.e.  less than 0.1ps, the curves for the
different system sizes fall on top of each other.  Thus for these short
times there are no finite size effects. This is not the case for longer
times, since we find that for $t>0.2$ps the curves for the different
system sizes show two types of size dependence (see inset of
Fig.~\ref{fig1}):  The first one is that the smaller the system is the
longer it takes the corresponding relaxation function to decay to
zero.  This effect is most pronounced for the smallest system, for
which we see that the relaxation is slower by more than a factor of 1.5
compared to the largest systems, but is also easily recognizable for
the system with $N=1002$. The second type of size dependence is that
the curves for the smaller systems show a local minimum at
$t=0.2$ps which is followed by an oscillatory behavior which is damped
out only for times larger than 2ps. For the larger systems this feature
is not observed at all, showing that it is a pure finite size effect.
Since this local minimum is believed to be related to the boson
peak~\cite{boson_simul,angell95}, care has to be taken for not
confusing the finite size effect with a real feature in the dynamics of
strong glass formers.

Although we present here only the results for the wave-vector
$q=2.8$\AA$^{-1}$ we have found a similar behavior also for different
values of $q$ (for 1.7\AA$^{-1} \leq q \leq 5.3$\AA$^{-1}$). In
particular we note that the time at which the blip is observed as well
as its amplitude is independent of $q$.

What is the reason for the occurrence of the oscillatory time
dependence in the correlation function for small systems? Lewis and
Wahnstr\"om have pointed out that a disturbance that propagates through
the system will leave and reenter the box due to the periodic boundary
conditions after a time of $L/c$, where $L$ is the size of the box and
$c$ is the typical velocity of the sound wave~\cite{lewis94}. This
disturbance is likely a sound wave with a small value of $q$ (since
modes with large values of $q$ are strongly damped). Because of
nonlinear effects this sound wave also couples to modes with larger
values of $q$ and thus gives rise to an echo for these values of $q$ as
well.  The returning wave thus gives rise to an additional signal in
the correlation function and hence the latter shows a slowed down
decay, as it is observed. The reason why this effect is so much
stronger in the system investigated here than in the molecular system
studied by Lewis and Wahnstr\"om is that in our system the damping of
the acoustic wave is significantly smaller than in their system,
probably due to the nature of the tetrahedral network in strong
glass formers as compared to the closed packed structure in fragile
glass formers.

In order to see how the magnitude of these finite size effects depend
on temperature we decreased the total energy per particle to a value of
$-18.6627$eV, which resulted in a temperature of around 1600K. The time
dependence of $F_{O}^{(s)}(q,t)$ for this lower temperature is shown in
Fig.~\ref{fig2}. We see again that for short times the curves for the
different system sizes coincide, whereas for larger times significant
differences occur. We see that now the local minimum at $t\approx
0.2$ps is present for all system sizes, thus indicating that this is
now a real feature of the dynamics of this system (and one that, as we
will show elsewhere~\cite{kob_boson}, is related to the boson peak).

For larger times the oscillations in the curves for the two smallest
systems are now much more pronounced, which shows that the damping of
the sound waves becomes smaller when the temperature is decreased.
Since therefore the magnitude of the signal that is added to the
correlation function, after the disturbance has crossed the box,
increases with decreasing temperature, the decay of the correlation
function is delayed more the lower the temperature is. In particular we
see that the difference in relaxation time between the system with
$N=336$ and the system with $N=8016$ is now around 4.0, thus
significantly larger than it was at the higher temperature.  (This
value was obtained by assuming that the main effect of the finite size
is to shift the curves horizontally, which is in accordance with the
observation in Fig.~1. A similar value is obtained from the mean
squared displacement, Fig.~3.) Such large differences are not tolerable
if one makes semiquantitative calculations in order to test whether or
not a given model is a good approximation to the real material.

Since we have shown that the magnitude of the finite size effects
depend on temperature it follows that if the temperature dependence of
the relaxation times are determined from a simulation of a small
system, this dependence will be different from the one for a large
system.  This means that also quantities like the glass transition
temperature, which is defined as the temperature at which a relaxation
time attains a certain value, will also depend on the system size.

In order to show that the observed effects occur not only in the
incoherent intermediate scattering function we have also investigated
the mean squared displacement (MSD) of a tagged particle, since this
quantity is commonly used to compute the diffusion constant. The
results for the oxygen atoms are shown in Fig.~\ref{fig3}.  As in the
case of the intermediate scattering function we find that the curves
for the different system sizes coincide for short times but show a
clear system size dependence for intermediate and large times. From
this figure we also recognize that in order to avoid finite size
effects it is necessary to use system sizes that are larger than 1000
particles.

Finally we come back to the point made above that the finite size
effects presented here are much weaker in fragile glass formers. In
order to test this hypothesis, we transformed the system
into a fragile glass former. This can be done by applying an external
pressure, since at sufficiently high pressure the tetrahedral network
of silica is destroyed and is replaced by a structure with a higher
coordination number which is more similar in nature to a simple liquid,
i.e.  a system of hard spheres or a Lennard-Jones fluid, than the
tetrahedral network is.  Since simple liquids are generally fragile or
intermediate glass formers~\cite{kob94}, it follows that compressing
silica will increase the fragility of the system and evidence for this
has indeed be found~\cite{boson_simul,angell95,barrat_96}.  Therefore
we increased the density of the system to 3.94g/cm$^3$ and repeated the
whole calculations for this new density.

In Fig.~\ref{fig4} we show the time dependence of $F_O^{(s)}(q,t)$ for
this new density for all system sizes investigated. The value of $q$ is
3.0\AA$^{-1}$, close to the location of the maximum in the O-O
structure factor at this density.  Two sets of curves are shown. The
first one is for a total energy per particle of -17.7343eV, which
corresponds to a temperature of about 3760K, the same temperature at
which we investigated the dynamics in the low density phase (see
Fig.~\ref{fig1}). The second set is for a total energy per particle of
-18.0586eV, and corresponds to a temperature of about 2860K.  At this
temperature the $\alpha$-relaxation time of the system is comparable to
the relaxation time of the system in the low density phase at
$T=3760$K. From this figure we see that the dynamics of the system is
now indeed very different from the dynamics in the low density phase,
in that the relaxation times are now much shorter. In addition we also
see that the system size dependence of the curves is now significantly
smaller than in the case at low density since the relaxation time for
the smallest system is now only 15\% larger than the one for the
largest system.  Also the temperature dependence of this remaining
finite size effect is now much weaker, if present at all, than it was
for the system at low density. Thus if we use the result that the high
density phase is indeed more fragile than the low density phase we can
conclude that the investigated finite size effects are much smaller in
fragile glass formers than in strong glass formers.

Before we conclude a few remarks are appropriate: i) Although we have
presented here only the results concerning the dynamics of oxygen, the
same conclusions also hold for the silicon atoms. ii) It has to be
emphasized that the finite size effects presented here do not at all
affect the results of simulations in which {\it static} quantities are
investigated. iii) Although we have investigated this sort of finite
size effect only for periodic boundary conditions, it can be expected
that they are also present for open boundary conditions or fixed
boundary conditions.  iv) We have also investigated whether structural
quantities, such as the radial distribution functions or the partial
structure factors show a dependence on system size and have found no
such dependence. This is in contrast with the results of Nakano {\it et
al.}, who found in their simulation of silica that the total structure
factor depends on the size of the system~\cite{nakano94}. The reason
for this discrepancy is that these authors compute the structure factor
from a Fourier transform of the radial distribution functions, which
leads to truncation effects, whereas we computed it directly from the
particle positions.

In this work we have shown that for a simple model for silica
surprisingly large finite size effects are present in the dynamics.
This result is likely to hold also for other strong glass formers and
therefore has to be taken into account in future computer simulations
of such systems.

Acknowledgments: This work was supported by BMBF Projekt 03 N 8008 C,
by SFB 262/D1 of the Deutsche Forschungsgemeinschaft and by the US
National Science Foundation under Solid State Chemistry Grant
DMR9108028-002.

\newpage

\begin{figure}
\caption{Time dependence of the incoherent intermediate scattering
function for oxygen for all system sizes investigated. $E=-18.0345$eV,
$T=3760$K, $\rho=2.33$g/cm$^{-3}$. Inset: Enlargement of the main figure. 
\protect\label{fig1}}
\vspace*{5mm}
\par

\caption{Time dependence of the incoherent intermediate scattering
function for oxygen for all system sizes investigated. 
$E=-18.6627$eV, $T=1600$K, $\rho=2.33$g/cm$^{-3}$. 
Inset: Enlargement of the main figure. 
\protect\label{fig2}}
\vspace*{5mm}
\par

\caption{Time dependence of the mean squared displacement
for oxygen for all system sizes investigated. $E=-18.6627$eV, 
$T=1600$K, $\rho=2.33$g/cm$^{-3}$. Inset: Enlargement of the main figure. 
\protect\label{fig3}}
\vspace*{5mm}
\par

\caption{Time dependence of the incoherent intermediate scattering
function for oxygen for all system sizes investigated.
$\rho=3.94$g/cm$^{-3}$. Lower set of curves: 
$E=-17.7343$eV, $T=3760$K. Upper set of
curves: $E=-18.0586$eV, $T=2860$K.
\protect\label{fig4}}
\vspace*{5mm}
\par

\end{figure}
\end{document}